\documentclass[preprint,12pt]{elsarticle}

\usepackage{lineno}

 \usepackage{graphicx, subfigure}
\usepackage{epsfig}
\usepackage{epstopdf}
\usepackage{hyperref}
\usepackage{caption} 
\usepackage{float} 
\usepackage{soul} 
 \usepackage{threeparttable}
\usepackage{amssymb}
\usepackage{amsmath}

\biboptions{sort}
\biboptions{compress}

\journal{Nucl. Instr. and Meth. in Phys. Res. A}

\begin{document}

\begin{frontmatter}



\title{Characterization and modeling of a low background HPGe detector}


\author[label1,label2]{N.~Dokania}
\author[label1,label2]{V.~Singh}
\author[label1,label2]{S.~Mathimalar}
\author[label3]{V.~Nanal\corref{cor1}}

\ead{nanal@tifr.res.in}
\cortext[cor1] {Corresponding author. Tel.: +91-22-22782333; fax: +91-22-22782133.}
\author[label3]{S.~Pal}
\author[label3]{R.G.~Pillay}
\address[label1]{India based Neutrino Observatory, Tata Institute of Fundamental Research, Mumbai 400 005, India.}
\address[label2]{Homi Bhabha National Institute, Anushaktinagar, Mumbai 400 094, India.}
\address[label3]{Department of Nuclear and Atomic Physics, Tata Institute of Fundamental Research, Mumbai 400 005, India.}
\begin{abstract}
A high efficiency, low background counting setup has been made at TIFR consisting of a special HPGe detector ($\sim$70$\%$) surrounded by a low activity copper+lead shield. Detailed measurements are performed with point and extended geometry sources to obtain a complete response of the detector. An effective model of the detector has been made with GEANT4 based Monte Carlo simulations which agrees with experimental data within 5$\%$. This setup will be used for qualification and selection of radio-pure materials to be used in a cryogenic bolometer for the study of Neutrinoless Double Beta Decay in $^{124}$Sn as well as for other rare event studies. Using this setup, radio-impurities in the rock sample from India-based Neutrino Observatory (INO) site have been estimated. 
\end{abstract}


\end{frontmatter}

\linenumbers
\section{Introduction}
\label{}
Understanding and minimization of background plays a very important role in rare decay studies like Double Beta Decay (DBD). For such rare processes (T$_{1/2} >$10$^{20}$ years), the sensitivity of measurement depends critically on the background level in the region of interest (ROI). The natural radioactivity from the surroundings ($\rm^{232}Th$ -- T$_{1/2}$ $\sim10^{10}$~years, $\rm^{235}U$ -- T$_{1/2}$ $\sim10^{8}$~years, $\rm^{238}U$ -- T$_{1/2}$ $\sim10^{9}$~years, $\rm^{40}K$ -- T$_{1/2}$ $\sim10^{9}$~years, etc.), setup materials and the detector itself are the source of $\rm\alpha, \beta, \gamma$ and neutrons. Further, muon-induced interactions in the materials surrounding the detector give rise to additional background of $\gamma$-rays and neutrons. While it is impossible to completely eliminate these background sources, it is essential to minimize the same. The flux of cosmic ray muons can be significantly reduced in an underground laboratory. Background from internal sources can be minimized by careful selection of radio pure materials~\cite{leonard, arnold}, while the background from the external sources is reduced by suitable shielding. In recent experiments, ultra low levels of background $\ge$10$^{-3}$ cts/(keV kg year) have been claimed using special materials and novel techniques \cite{agostiniprl, auger}. The total background, both from external and internal sources, has to be taken into consideration during the interpretation of results. Generally, a background model employing Monte Carlo (MC) simulations taking into account all the contributions from the actual setup and the environment in the experimental site is used for physics analysis \cite{agostini, bellini, andreotti, argyriades}. 

To assess the level of radio purity in the materials surrounding the detector, samples are often counted in a close geometry to obtain high counting efficiency. For accurate determination of radio impurities, precise knowledge of detection efficiency over a wide energy range is necessary. 
The efficiency measurement in a close geometry is complicated using standard multi-gamma sources due to coincidence summing effects. Hence, measurements are restricted to available mono-energetic sources in a limited energy range. Consequently, MC simulation technique is adopted to obtain efficiency of the detector over a wide energy range for different source-detector configurations. It has been observed in the literature \cite{budjas, cabal, diaz, hardy, helmer, hernandez, hurtado, karamanis2003} that the efficiency computed from the MC simulations using the detector geometry supplied by the manufacturer is overestimated (by $\geq$10$\%$) as compared to the experimental values. The discrepancy in efficiency is attributed to the inaccuracy of the supplied parameters, like detector size and the dead layer. It should be mentioned that this effect is more pronounced for large size detectors \cite{helmer, hernandez}, which may be due to incomplete charge collection. Thus, the parameters of the detector need to be optimized by detailed measurements along the detector surfaces covering the energy range of interest. In addition, measurements at different distance for various source geometries are required to extract the active volume. 

A feasibility study to search for 0$\nu\beta\beta$ in $^{124}$Sn using a tin cryogenic bolometer \cite{Raina, VSingh, pramana} has been initiated at the upcoming underground facility in India-based Neutrino Observatory (INO) \cite{ino}. In case of $^{124}$Sn, $\rm Q_{\beta\beta}$=2.293 MeV \cite{qvalue} is close to the Compton edge of 2.614~MeV $\gamma$--ray, originating in the decay chain of $\rm^{232}Th$ ($\rm^{208}Tl {\xrightarrow{\beta^-}} {^{208}Pb (3^-)} {\xrightarrow{\gamma}} {^{208}Pb (0^+)}$). To investigate the background issues pertaining to NDBD search in $^{124}$Sn, a low background counting setup with HPGe detector has been made at sea level in TIFR, Mumbai. 
This setup is intended for screening of materials in the prototype bolometer R$\&$D at TIFR as well as for understanding the background. In addition, the setup will be used for rare event studies like DBD to the excited states of daughter nuclei, rare alpha decays etc.~\cite{barabash2, belli, belli1, belli2, barabash3, belli3}. This paper describes the optimization of the HPGe detector model using MC simulations. In the present work, mono-energetic sources are used to scan the Ge crystal in directions parallel and perpendicular to its cylindrical axis. Measurements are also done with sources over an energy range of E$_{\gamma}$=100-1500~keV as a function of distance to estimate its active volume. Experimental details are discussed in Section 2. Section 3 describes the procedure of MC simulations as well as the optimization of different parameters of the crystal, namely, the top and side dead layer, front gap, radius, length and hole size. Results of the detector model and measurements in the low background counting setup are presented in Section 4 and conclusions are given in Section 5.

\section{Experimental Details}
\begin{figure}[!h]
        \begin{minipage}[b]{0.45\linewidth}
          \centering
          \includegraphics[scale=0.9]{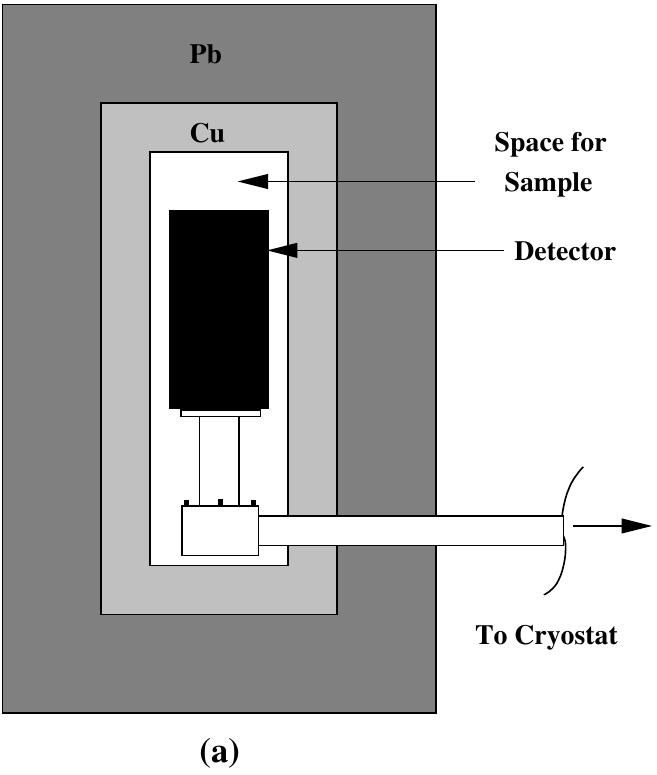}
        \end{minipage}
        \hspace{0.4cm}
        \begin{minipage}[b]{0.45\linewidth}
          \centering
          \includegraphics[scale=0.3]{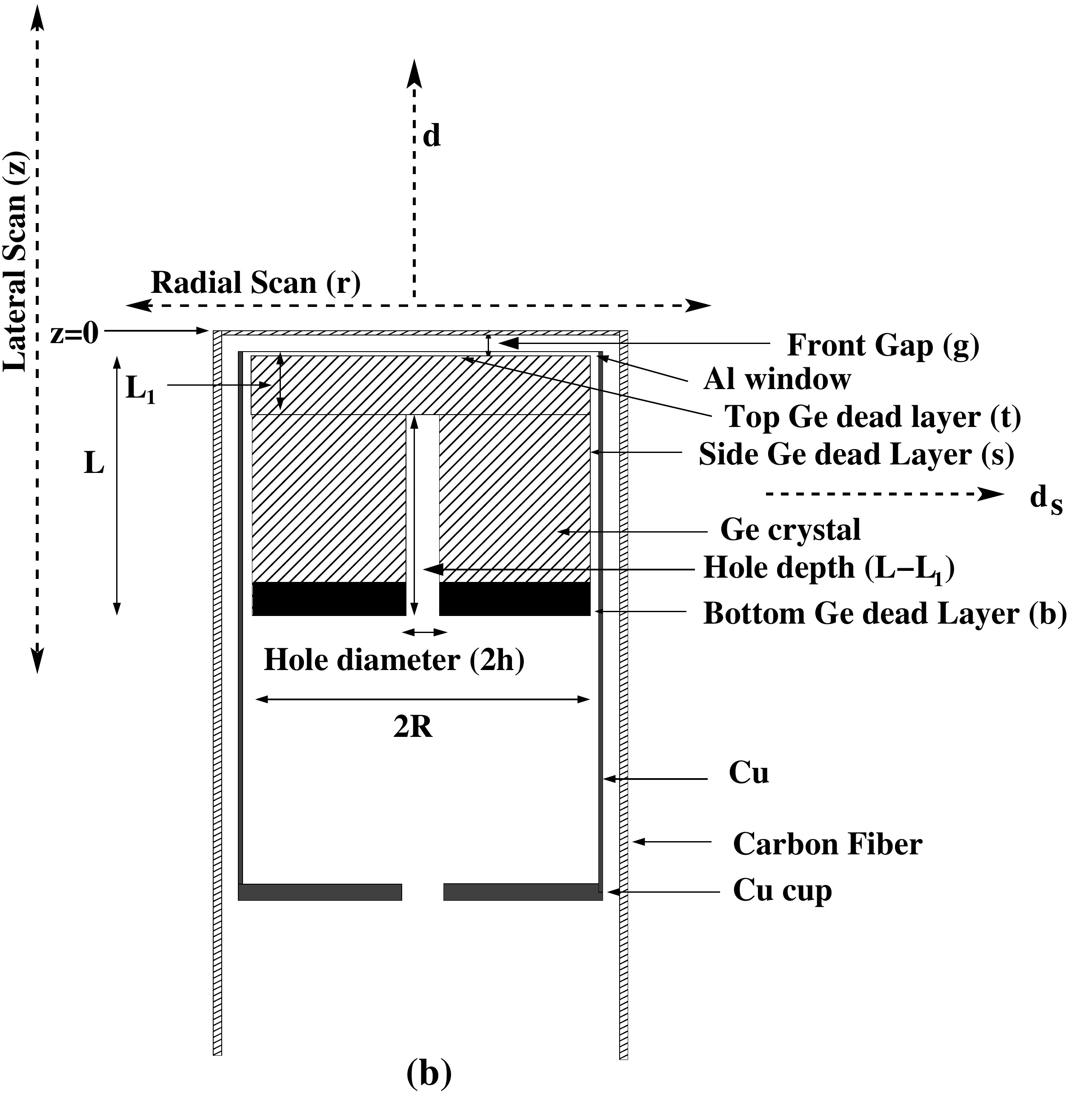}
        \end{minipage}
        \caption{\label{sketch} (a) A schematic view of low background counting setup comprising the HPGe detector, $\sim$5~cm Cu shield, outer 10~cm low activity Pb shield and showing the space for sample, (b) A cross-sectional view of the detector showing different parameters. Scan directions for lateral, radial, top ($d$) and side ($d_S$) are also indicated. The center of the detector corresponds to $r$=0 and the top edge of the detector, i.e. carbon fiber housing, corresponds to $z$=0.}
      \end{figure}

The HPGe detector is a coaxial p-type Ge (ORTEC GEM75-95-LB-C-HJ), specially designed for low background measurements with a relative efficiency of $\sim$70$\%$. It has a low background carbon fiber outer body and copper support structures with a 60~cm long cold finger attached to a J-shaped cryostat. Figure~\ref{sketch}a shows a schematic view of the experimental setup together with the inner 5~cm low activity OFHC Cu shield and the outer 10~cm low activity Pb shield ($<$0.3~Bq/kg $\rm^{210}Pb$). Figure~\ref{sketch}b shows the cross-sectional view of the detector indicating different parameters. The detector bias used is +4~kV, as recommended by the manufacturer. The nominal size of the Ge crystal given by the manufacturer is 78.3~mm diameter and 63~mm length with a 0.7~mm dead layer on the cylindrical side. In addition to electrical contacts, the detector is surrounded by aluminized mylar and thin copper on sides as well as on bottom for thermal shielding. Generally, the physical dimensions of the detector can be determined by radiography \cite{budjas, boson} but the active volume of the detector may differ depending on the electric field configuration inside the crystal \cite{hernandez}. Precise measurements of photopeak efficiencies using radioactive sources give better estimates on the actual active volume and the surrounding materials of the detector. In the present case, radiography of the setup was not possible and hence mono-energetic sources covering an energy range of 59.5--1115.5~keV were used to scan the crystal. 
Table~\ref{table1} gives the details of various sources and source geometries used in the present work together with respective gamma ray energies. 

\begin{figure}[H]
\centering
\includegraphics[scale=0.5]{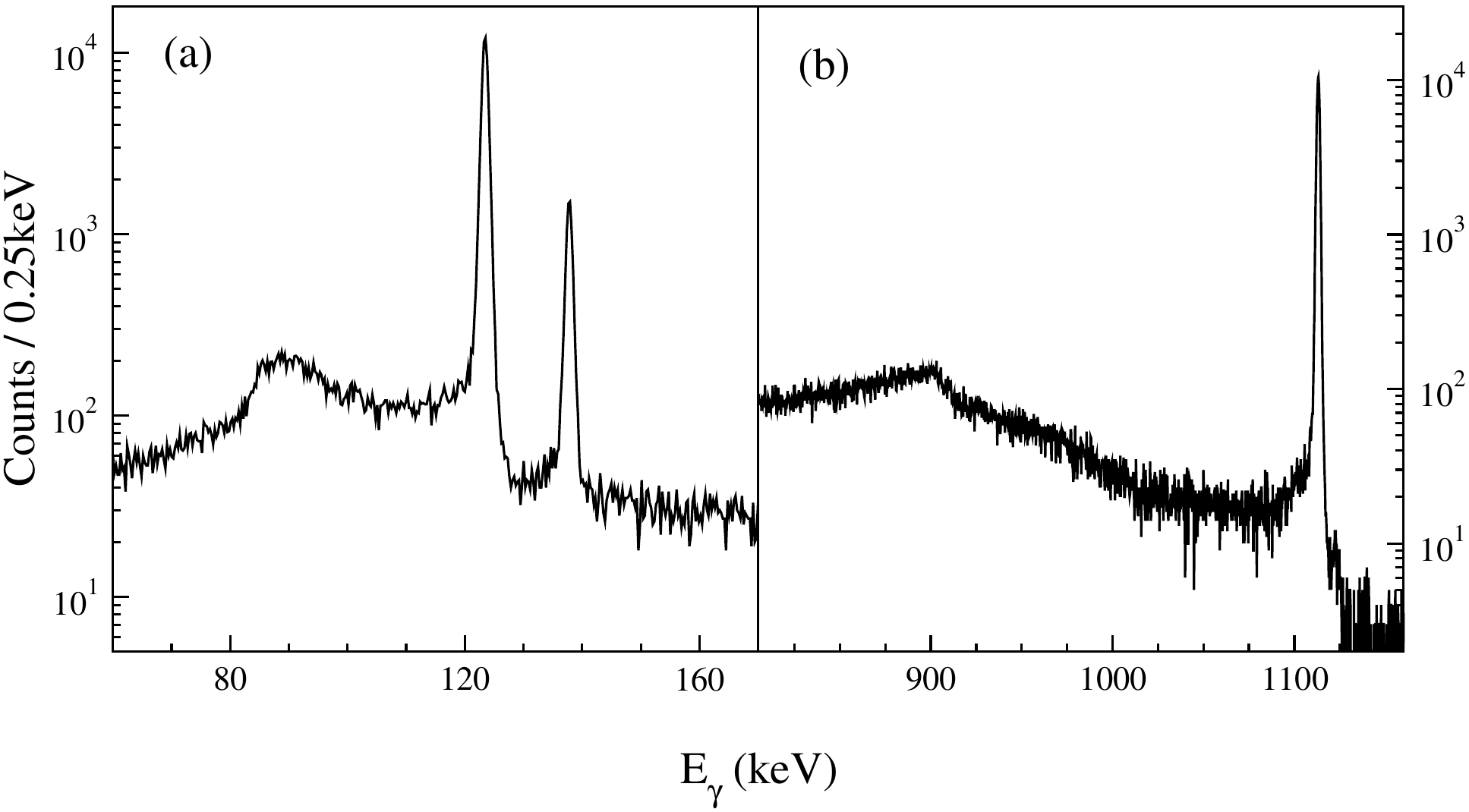}
\caption{\label{res} Gamma ray spectra obtained with (a) $\rm^{57}Co$ at $d$=10~cm, and (b) $\rm^{65}Zn$ source at $d$=1~cm. }
\end{figure}

\begin{table}[H]
\centering
\caption{List of radioactive sources used for measurements.}
\vspace{0.6em}
\label{table1}
\begin{tabular}{  ccc }
\hline
Isotopes& Energy& Geometry \\ 
\it & (keV)& \\ \hline
$\rm^{241}Am$ & 59.5 & point \\ 
$\rm^{57}Co$  & 122.1 & extended\\ 
$\rm^{203}Hg$  & 279.2 & extended\\
$\rm^{51}Cr$  & 320.1 & extended\\ 
$\rm^{137}Cs$  & 661.7 & volume\\ 
$\rm^{54}Mn$  & 834.8 & extended\\ 
$\rm^{65}Zn$  & 1115.5 &extended\\ \hline
$\rm^{152}Eu$  &121.8, 778.9, 1408& point\\ 
$\rm^{60}Co$ & 1173.2, 1332.5&point \\ \hline
\end{tabular}
\end{table}
Measured absolute strengths of sources are in the range of $\sim$1-90~kBq with $\sim$0.8-1.5$\%$ uncertainty. The extended geometry source has a 6~mm active diameter and is mounted on a 25~mm diameter plastic disc with a 1~mm thick plastic front cover. In case of $\rm^{137}Cs$ volume source, the liquid was sealed inside a perspex cylindrical vial of radius 3~mm and height 5~mm. The distribution of $\rm^{137}Cs$ volume source was assumed to be homogeneous in the perspex vial. Measurements for optimizing detector geometry can be broadly classified into three categories (see Figure~\ref{sketch}b), namely, radial scan, lateral scan and distance scan for volume effect. Radial and lateral scans are carried out with $^{241}$Am, $^{57}$Co and $^{65}$Zn sources. The low energy gamma-rays are sensitive to the dead layers and high energy gamma-rays probe the detector size. Radial scan was done by moving the source parallel to the top detector face ($r$) at a distance of 5~mm in 3~mm steps and covered a range of $\pm$6~cm w.r.t. the center of the detector. 
For the lateral scan the source was moved parallel to its cylindrical axis ($z$) at a distance of 8~mm from the side face of the detector in 3~mm steps and covered a range of $\pm$8~cm w.r.t. the top face of the detector. The distance scan ($d$) was done in steps of 5~cm over a distance of 0--25~cm from the top face as well as from the cylindrical side of the detector to study the volume effect for E$_{\gamma}$=834.8 and 1115.5~keV. Typical uncertainty in positioning of the source, both in horizontal and vertical direction, was less than 1~mm. Detector signal was given to a 13-bit analog-to-digital converter through a spectroscopic amplifier (shaping time : 10$\mu$s). Data was recorded with a CAMAC-based acquisition system, LAMPS \cite{lamps}. Dead time correction was done using a standard 10~Hz pulser. Figure~\ref{res} shows gamma-ray energy spectra with {$^{57}$Co} and {$^{65}$Zn}. Typical measured energy resolution (FWHM) obtained was 0.75(2)~keV at 122.1~keV and 1.84(2)~keV at 1115.5~keV, respectively. Photopeak efficiency ($\rm\epsilon^{exp}$) was extracted using LAMPS software by fitting the observed photopeak to a Gaussian function with either a linear or a quadratic background. In some cases, the observed peak had a slight low energy tail, which could be incorporated in the fitting software. However, the contribution from tail region was found to be negligible. In the present case, given relatively low source strengths no pile up effects have been observed in the spectra. Errors were computed including statistical errors and least-squares fitting errors in extracting the peak areas. Typical errors obtained in $\rm\epsilon^{exp}$ were : in radial/lateral scans $\sim$3.7$\%$ for $\rm E_{\gamma}$=59.5~keV, 0.2$\%$ for $\rm E_{\gamma}$=122.1~keV and 1.8$\%$ for $\rm E_{\gamma}$=1115.5~keV. It should be mentioned that differences in statistical errors are mainly due to the difference in strengths of various sources and energy dependent variation in detection efficiency. Similarly, for both the top and side distance scan errors in $\rm\epsilon^{exp}$ were $\sim$2$\%$ and $\sim$5$\%$, respectively. 

To verify the detector model, additional radial scans with $^{57}$Co and $^{65}$Zn sources at $d$=10.7~cm were carried out and distance scan ($\sim$1--30 cm) was done with various sources covering an energy range of 122.1--1408~keV. In this case, multi-gamma sources such as $\rm^{152}Eu$ and $\rm^{60}Co$ were used at a distance $d>$10~cm to ensure that the coincidence summing is negligible. Measurements were also done with the volume source ($\rm E_{\gamma}$=661.7~keV).

\section{Monte Carlo Simulations}
In the present work, GEANT4 (version 4.9.5.p01) \cite{geant} is used to simulate the HPGe detector response. 
The coaxial geometry of Ge crystal with a central hole is realized in the simulations by placing a circular disk of radius $R$ and thickness $L_1$ on a hollow cylinder of length $L-L_1$, as shown in Figure~\ref{sketch}b. 
The inner radius of the hollow cylinder is taken to be that of the hole ($h$) and the outer radius is $R$. The curvature of the edges of the cylinder/disk is neglected. 
Complete details of the surrounding absorbing materials such as top and side Ge dead layers, Al window, Cu cup support structures, outer carbon fiber body have been included in the Monte Carlo model.
 Source geometry is also taken into account in the MC simulations.
 It should be mentioned that the MC code is verified with other HPGe detector geometries \cite{cabal, hurtado}. A photon of given energy is generated in the MC simulations.
Simulations have been carried out for a set of detector parameters over a range of $r$ and $z$ in 6~mm steps corresponding to the measurements. 
 Event by event data obtained from MC is binned in 0.25~keV bin size and absolute photopeak efficiency ($\rm\epsilon^{MC}$) is determined using the ROOT analysis framework \cite{root}. 
In some cases where the source co-ordinates in the experiment ($r_i$, $z_i$) were different from those in the simulation (diff.$\sim$1~mm), the $\rm\epsilon^{MC}$ corresponding to $r_i$, $z_i$ was obtained by interpolation. 
Statistical uncertainties are kept below 2$\%$. 
For modeling the detector geometry, only absolute photopeak efficiencies of different $\gamma$--rays are taken into consideration. 
The best fit values of detector parameters are obtained by two methods. 
In the first method, $\rm{{\chi}^2}$ is determined for a data set like radial/lateral/distance scan ($n$ points) corresponding to each source~\cite{knoll} using Eq.~\ref{chi},

\begin{equation}\label{chi}
\rm{{\chi}^2} =\frac{1}{n-1}\rm{\sum_{i=1}^{n}{\frac{ (\epsilon_{E}^{exp}[r_i] - \epsilon_{E}^{MC}[r_i])^2}{{\epsilon_{E}^{MC}[r_i]}}} }
\end{equation}
where, $\epsilon^{exp}_E(r_i)$ represents the measured absolute photopeak efficiency at $\rm r_i$ for a $\gamma$--ray of energy E$_{\gamma}$ and $\epsilon^{MC}_E (r_i)$ is the corresponding simulated efficiency. In the second method, following the procedure as in \cite{helmer, hardy} to give similar weightage to $\epsilon_E$ for different energies, the total relative deviation between measured and simulated efficiencies is calculated as defined in Eq.~\ref{totalrd},

\begin{equation}\label{totalrd}
\rm{{\sigma}{_R}} =\frac{1}{n_{2}}\sum_{j=1}^{n_2}\left\{\frac{1}{n_{1}}\rm{\sum_{i=1}^{n_1}{\frac{ \mid{\epsilon_{E_j}^{exp}[r_i] - \epsilon^{MC}_{E_j}[r_i]}\mid{}}{{\epsilon_{E_j}^{MC}[r_i]}}} }\right\}
\end{equation}
 where $n_1$ is number of points in each data set and $n_2$ is number of data sets corresponding to different energies or scans.

\subsection{Optimization of detector model}
 It is observed from the simulation data that the measured value of 66$\%$ relative efficiency corresponds to an active volume of $\sim$230~cm$^3$, which is significantly smaller ($\sim$20$\%$) than the number quoted by the manufacturer (292~cm$^3$). 
Further, a comparison of $\epsilon_E^{MC}$ using default detector parameters with $\epsilon_{E}^{exp}$ for E$_{\gamma}$=122.1 to 1115.5~keV and $d$=5 to 25~cm, resulted in a large relative deviation, $\rm\sigma_R\sim$29.2(3)$\%$.
The response of the central core region of the detector was probed by measurements with two collimators made from a 5~cm thick lead block with a 13~mm (35~mm) diameter conical (cylindrical) hole at the center.
In both cases, a better agreement has been observed between the simulations and the measured values for the restricted central volume of the detector. 
It is therefore necessary to optimize the size of the detector to reproduce the experimental data. 
For generating the detector model, the crystal parameters varied are (see Figure~\ref{sketch}b) : top Ge dead layer ({\it t}), side Ge dead layer ({\it s}), front gap ({\it g}) i.e., the distance between the top carbon fiber and the Al window, crystal radius ({$R$}), crystal length ({$L_1$ and $L$}) and hole radius ($h$). 
External detector parameters like thicknesses of carbon fiber housing, Al window and Cu cup are taken as given by the manufacturer.
Initial crystal parameters, namely, radius ($R_i$=37.5~mm), length ($L_i$=55~mm), hole radius ($h_i$=6.5~mm) and front gap ($g_i$=5~mm) were obtained by the best fit to the scan data of E$_{\gamma}$= 1115.5~keV at close distance, where measurements are not strongly affected by the dead layers and surrounding materials. 
For the front gap estimation, the fit has been restricted to the central region i.e. $r$=$\pm$3~cm, to minimize the effect of radial extension of crystal.
 
The dead layer on the crystal attenuates the gamma rays and is best estimated with low energy gamma rays. 
It reduces the active volume of the detector \cite{rodenas} and may also increase with time depending on years of operation \cite{huy}. 
As mentioned earlier, no top dead layer ($t$) has been specified by the manufacturer while the side dead layer ($s$) is quoted as 0.7~mm. 
The uniform dead layer is employed in the simulations and values of $t$ and $s$ are varied in the range of 0--1.2~mm and 0.7--1.5~mm, respectively. 
It should be mentioned that a $2\%$ variation in dead layer thickness results in $\sim$ 2$\%$ change in the photopeak efficiency for E$_{\gamma}$=59.5~keV.
The $\sigma_R$ is calculated for the central region of radial (lateral) scan, namely, $r$=$\pm$3~cm ($z$=$\pm$2.5~cm), with 59.5~keV and 122.1~keV $\gamma$--ray sources mounted close to the face of the detector.   
The best fit values of $t$ and $s$ extracted corresponding to a minimum $\sigma_R$ are $t_{opt}$=1.04$\pm$0.02~mm and $s_{opt}$=1.27$\pm$0.02~mm. 

The germanium disc thickness $L_1$ was obtained by fitting the $\epsilon^{exp} (r=0)$ data of E$_{\gamma}$=320.1~keV close to the detector top face. 
Since for this energy halfvalue layer for germanium is $\sim5$~mm, the $\epsilon^{MC}$ is expected to have better sensitivity for $L_1$ and has a very little dependence on dead layers. 
The $L_1$ was varied from 7.5~mm to 12.3~mm in steps of 0.2~mm and minimum $\chi^2$ was found at $L_{1-opt}$=9.7$\pm$0.5~mm.
Considering the physical length specified by the manufacturer ($L_m$), an inactive Ge dead layer of thickness $b$=$L_m - L$ surrounded by a 3.5$\pm$0.5~mm thick cylindrical Cu ring at the bottom of the crystal is included in the model. 
This resulted in a better reproduction of the overall shape of the measured lateral scan for low energy gamma-rays.

\begin{figure}[H]
\centering
\includegraphics[scale=0.5]{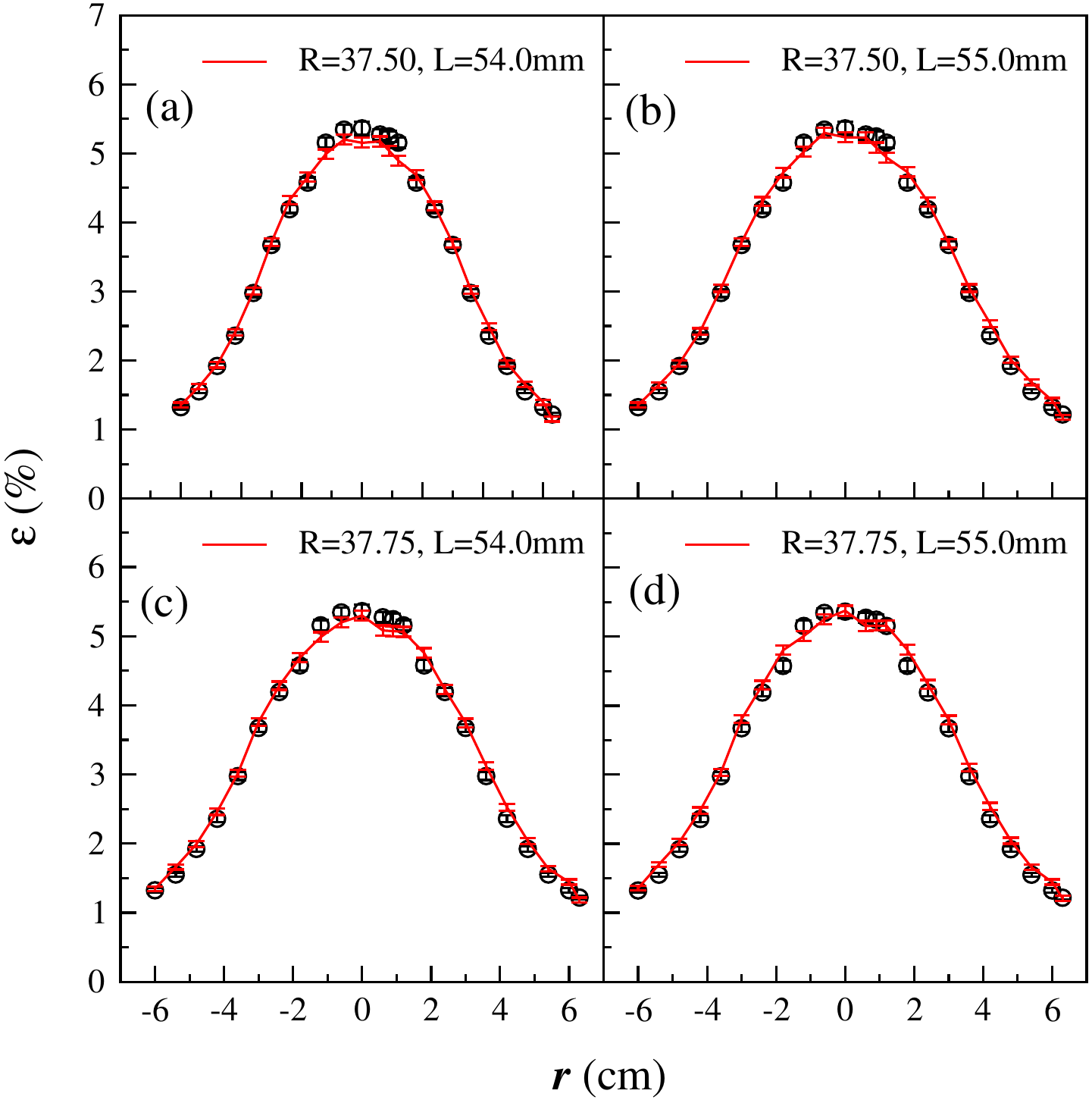}
\caption{\label{zn_rad}(Color online) The absolute photopeak efficiency $\epsilon^{exp}$ (unfilled circles) of E$_\gamma$=1115.5~keV as a function of $r$ (radial scan). The simulated values $\epsilon^{MC}$ (lines) for different combinations of radii ($R$) and lengths ($L$) are shown in panels (a) to (d).}
\end{figure}

\begin{figure}[H]
\centering
\includegraphics[scale=0.5]{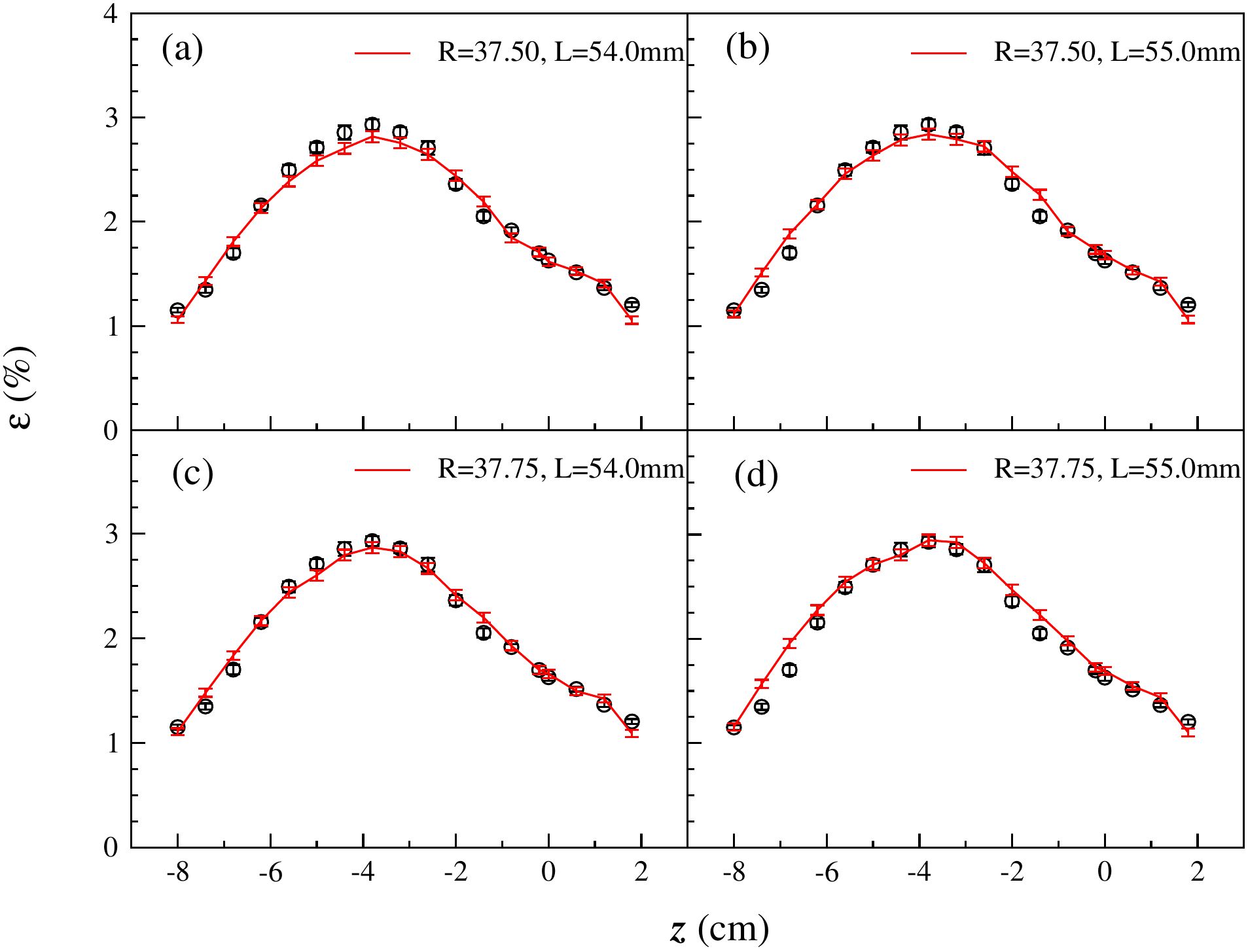}
\caption{\label{zn_length}(Color online) The absolute photopeak efficiency $\epsilon^{exp}$ (unfilled circles) of E$_\gamma$=1115.5~keV as a function of $z$ (lateral scan). The simulated values $\epsilon^{MC}$ (lines) for different combinations of radii ($R$) and lengths ($L$) are shown in panels (a) to (d).}
\end{figure}

For extracting $R_{opt}$ and $L_{opt}$, simulations have been carried out by varying $R$ and $L$ in fine steps of 0.25~mm and 1~mm, respectively. 
Figures~\ref{zn_rad} and \ref{zn_length} show $\epsilon^{exp}$ together with $\epsilon^{MC}$ for the radial and lateral scan of E$_{\gamma}$=1115.5~keV, respectively. 
 It is evident that $R$ and $L$ are not independent of each other. Therefore, the best fit values of $R$ and $L$ are obtained by a simultaneous fit to the radial and the lateral scan data for E$_{\gamma}$=1115.5~keV.
 
\begin{figure}[H]
\centering
\includegraphics[scale=0.5]{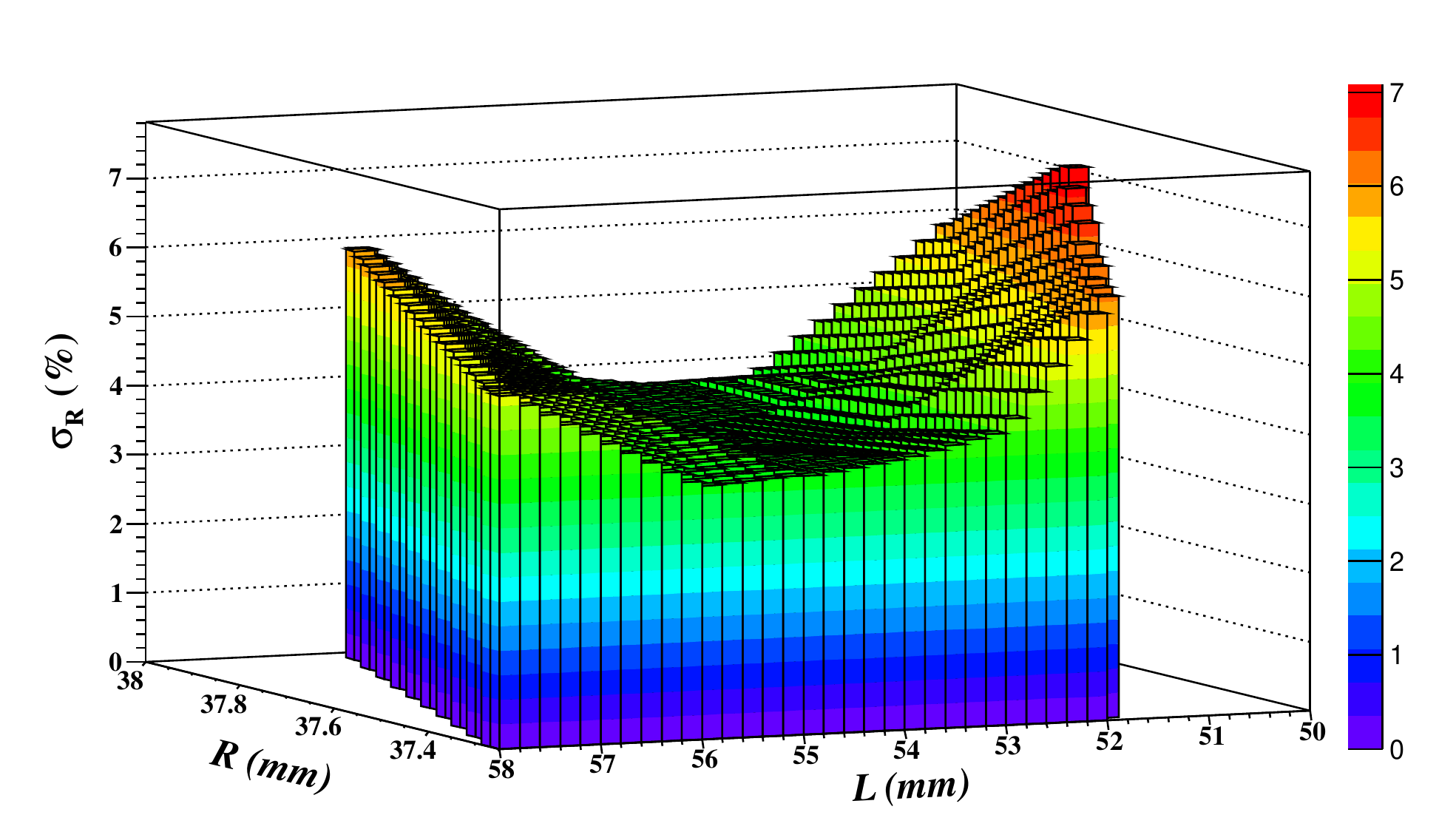}
\caption{\label{sigma}(Color online) The total relative deviation $\sigma_{R}$ as a function of $R$ and $L$ for both radial and lateral scans with E$_\gamma$=1115.5~keV.}
\end{figure}

Figure~\ref{sigma} shows a pictorial representation of the $\sigma_R$ for radial and lateral scan. It can be seen that the minimum is rather shallow. The $R_{opt}$ and $L_{opt}$ are obtained from a weighted mean over the region of the shallow minimum in $R$-$L$ space with weight for each point taken as $\sigma_R^{-1}$. The optimal values obtained after rounding off to the first decimal place are $R_{opt}$=37.6$\pm$0.3~mm and $L_{opt}$=54.0$\pm$0.9 mm. The errors quoted are the standard deviations on the calculated quantities.

With above values of $R_{opt}$ and $L_{opt}$, the hole depth $L_h$=44.3 $\pm$1.0~mm was obtained corresponding to $L_{opt}$--$L_1$. 
The remaining unknown parameter, hole radius $h$, was extracted from the distance scan with high energy $\gamma$--rays. 
From the fit of the distance scan data (1--25 cm) of E$_\gamma$= 834.8 and 1115.5~keV, the $h_{opt}$ was found to be 7.5$\pm$0.6~mm.
As mentioned earlier, the bottom dead layer $b_{opt}$ was set to the difference between $L_m$ and $L_{opt}$. 
Table~\ref{final} gives a complete list of optimized parameters of the detector. Errors on the parameters have been estimated from the standard deviations on the calculated quantities. The quantities marked with an asterisk in Table~\ref{final} have not been altered in the MC simulations.
The nominal parameters supplied by the manufacturer are also shown for the comparison.

\begin{table}[H]
\centering
\begin{threeparttable}[b]
\caption{Optimized parameters of the detector.}
\label{final}
\begin{tabular}{  ccc }
\hline
Detector& Nominal& Optimized \\ 
Parameter  &(mm) &(mm) \\ \hline
Ge crystal radius ($R$) & 38.45 & 37.6$\pm$0.3\\ 
Ge crystal total length ($L$) & 63.0    & 54.0$\pm$0.9\\ 
Ge disc thickness ($L_1$) & 12.3    & 9.7$\pm$0.5\\ 
Hole depth ($L-L_1$) & 50.7    &44.3$\pm$1.0\\
Hole radius ($h$)         & 5.5   &7.5$\pm$0.6\\
Top Ge Dead Layer ($t$)    & -     & 1.04$\pm$0.02\\ 
Side Ge Dead Layer ($s$)   & 0.7   & 1.26$\pm$0.02\\ 
Bottom Ge Dead Layer ($b$)   & -   & 9.0$\pm$1.0\\ 
Front gap ($g$) & 4 & 5.0$\pm$0.7\\
Top carbon fiber\tnote{*}&0.9&0.9\\
Side carbon fiber\tnote{*}&1.8&1.8\\
Cu Cup thickness\tnote{*}&0.8&0.8\\
Ge Crystal Volume ($V$)&292~$\rm cm^3$&232$\pm$6~$\rm cm^3$\\\hline
\end{tabular}
\begin{tablenotes}
\item[*]Not altered in MC simulations
\end{tablenotes}
\end{threeparttable}
\end{table}

\section{Results}
\subsection{Validation of detector Model}

\begin{figure}[H]
        \begin{minipage}[b]{0.45\linewidth}
          \centering
          \includegraphics[scale=0.6]{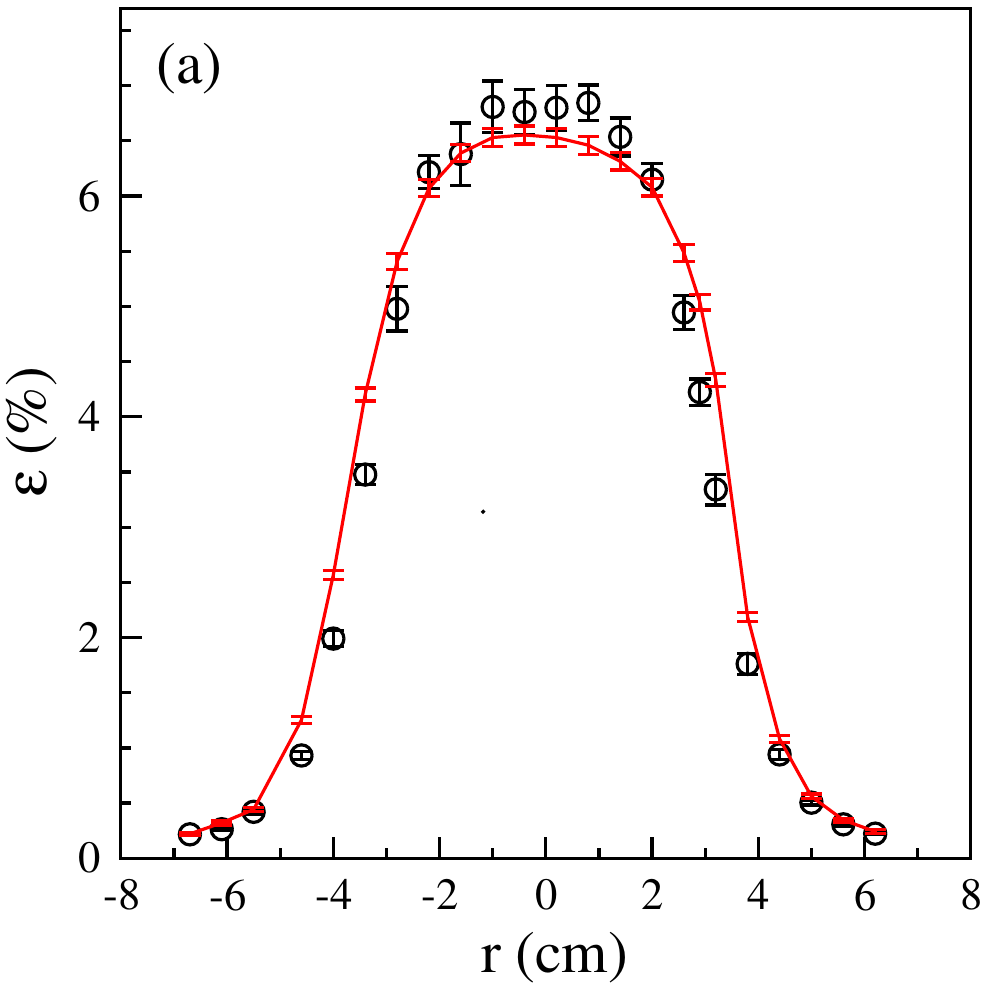}
          \end{minipage}
        \hspace{0.2cm}
        \begin{minipage}[b]{0.45\linewidth}
          \centering
          \includegraphics[scale=0.6]{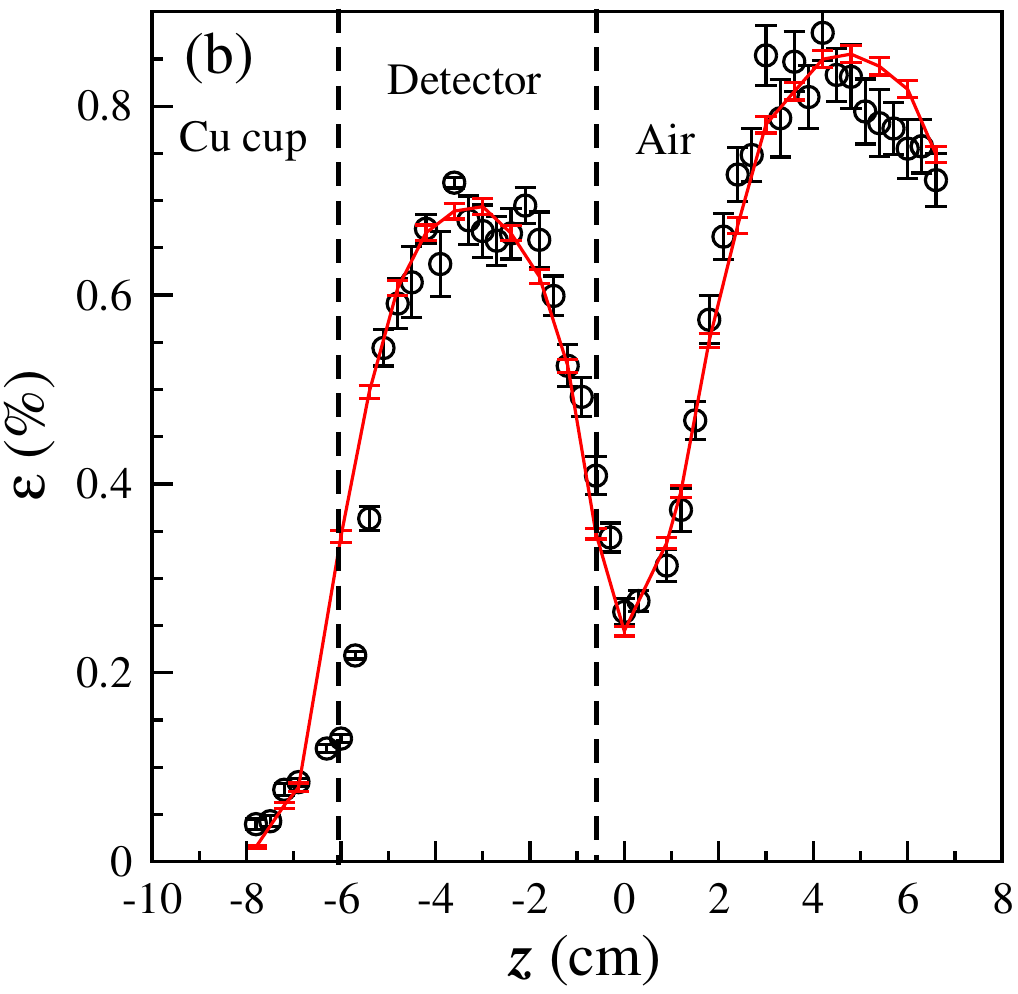}
               \end{minipage}
        \caption{\label{sd59}(Color online) The radial (left panel) and lateral (right panel) scan data of E$_{\gamma}$=59.5~keV with optimized detector parameters. Symbols represent the $\epsilon^{exp}$ and the line corresponds to $\epsilon^{MC}$. The $z$ range occupied by the crystal is marked in the figure. }
      \end{figure}

\begin{figure}[H]
\centering
\includegraphics[scale=0.6]{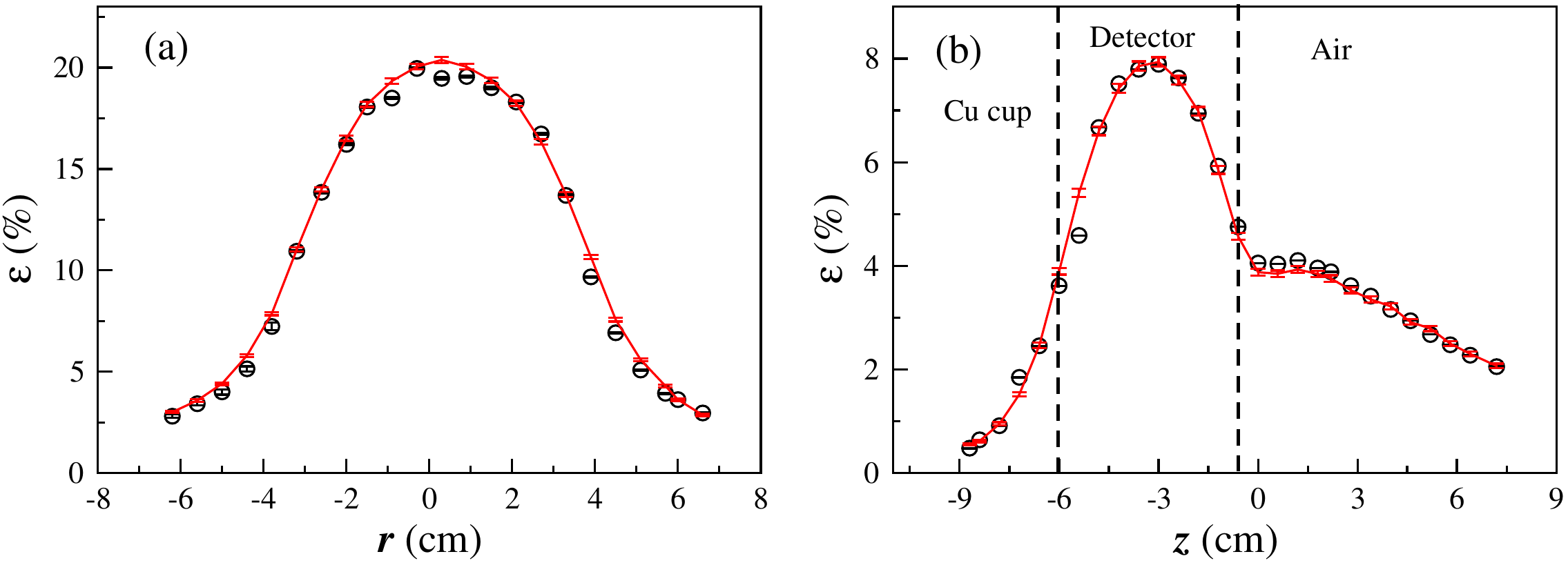}
\caption{\label{sd122}(Color online) Same as Figure~\ref{sd59} for E$_{\gamma}$=122.1~keV }
\end{figure}

\begin{figure}[H]
\centering
\includegraphics[scale=0.6]{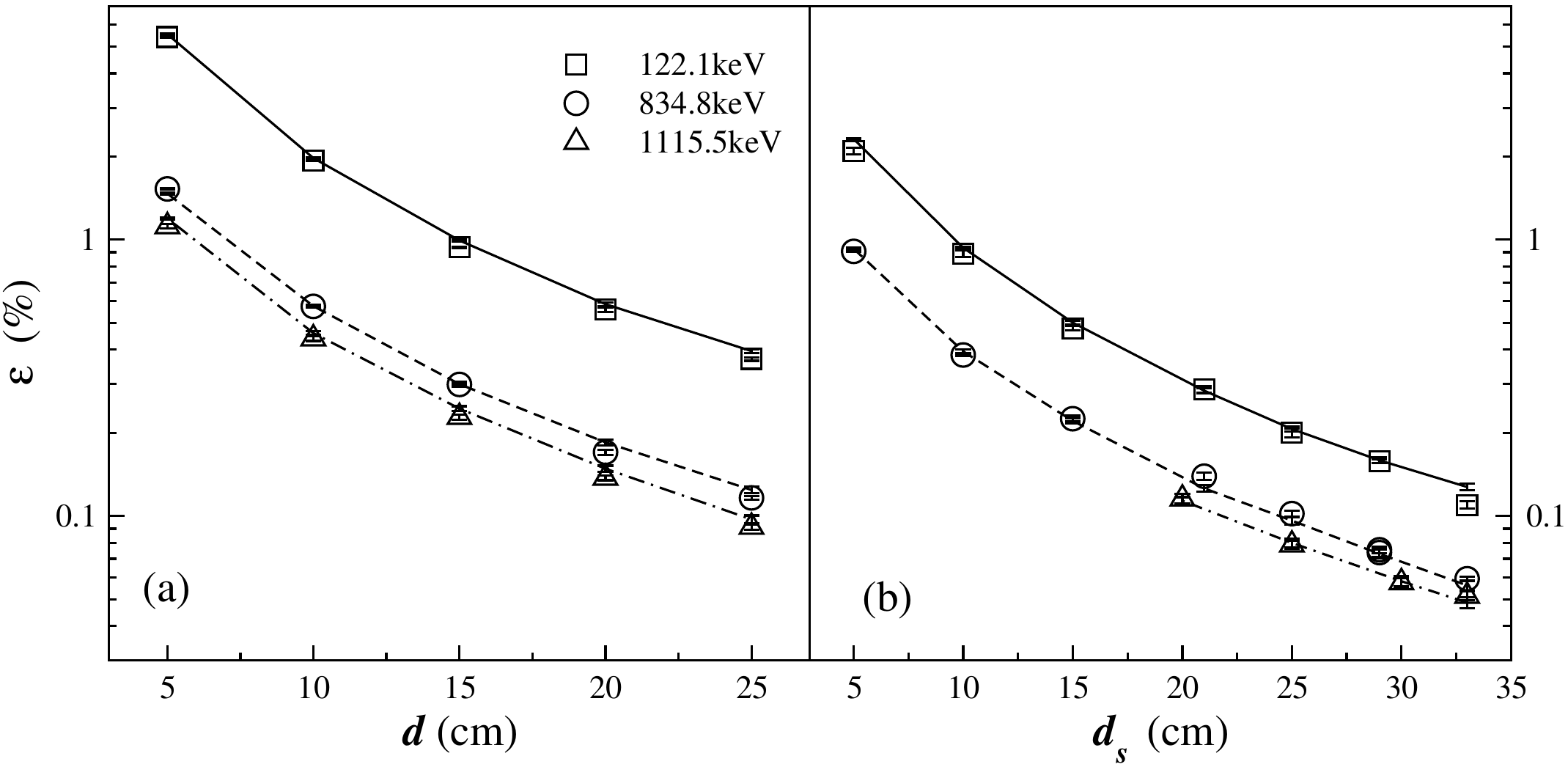}
\caption{\label{d834} The top distance (left panel) and side distance (right panel) scan data of E$_{\gamma}$=122.1,~834.8~and~1115.5~keV with optimized detector parameters. Symbols represent the $\epsilon^{exp}$ and the line corresponds to $\epsilon^{MC}$. }
\end{figure}

\begin{figure}[H]
\centering
\includegraphics[scale=0.6]{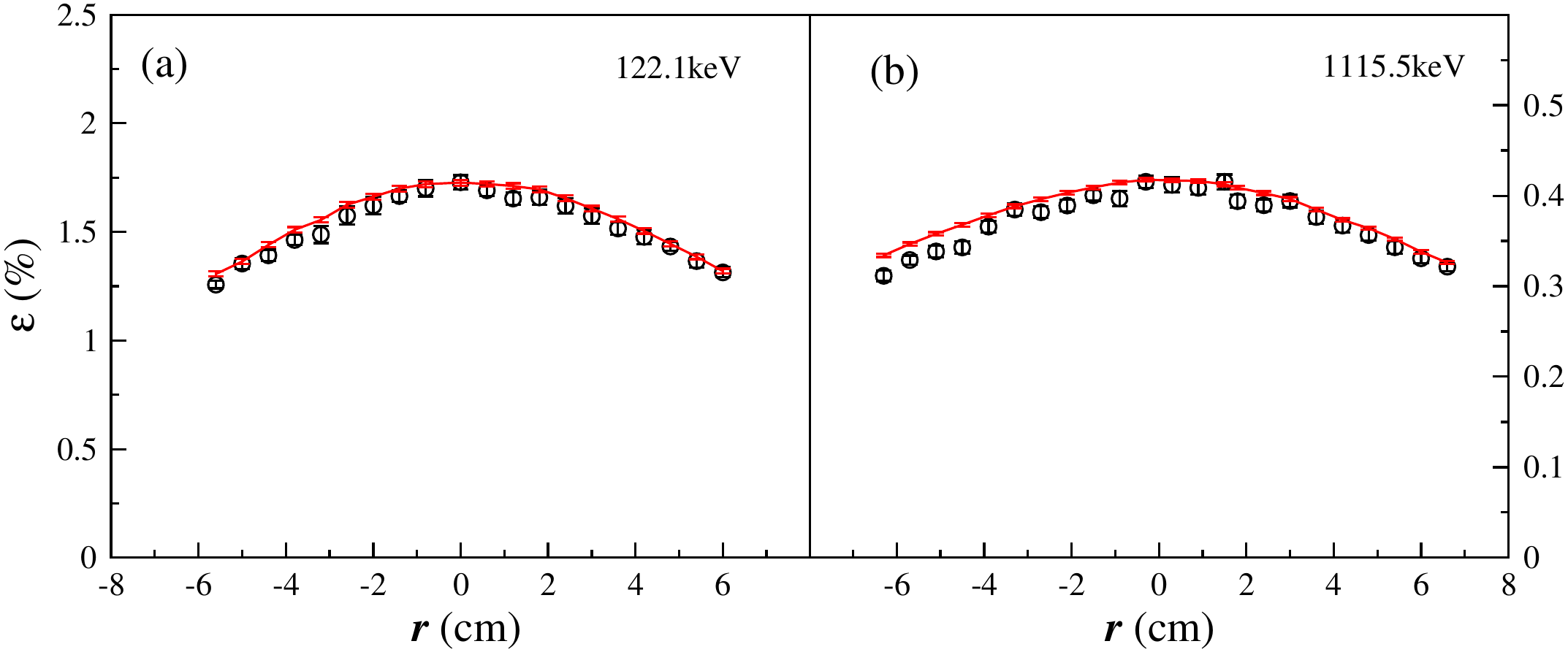} 
\caption{\label{rad10}(Color online) The radial scans data showing $\epsilon^{exp}$ (unfilled circles) and $\epsilon^{MC}$ (lines) for (a) E$_{\gamma}$=122.1~keV and (b) E$_{\gamma}$=1115.5~keV with optimized detector parameters at $d$=10.7~cm.}
\end{figure}

Figures~\ref{sd59},~\ref{sd122} and~\ref{d834} show a comparison of experimental data for various energies together with simulation results employing the optimized detector parameters. For the lateral scan with low energy gamma-rays, addition of the bottom dead layer ($L_m-L_{opt}=b$) is crucial to reproduce the shape in $z$=--9 to --6~cm region (see Figures~\ref{sd59}b and~\ref{sd122}b). It should be mentioned that at low energy (E$_{\gamma}$=122.1~keV), the effective linear dimension of the crystal (radius/length) seems to be lower than that for the high energy (E$_{\gamma}$=1115.5~keV). 
This could be an effect of a non-uniform electric field at corners of the crystal~\cite{hernandez} or the non-uniform dead layer~\cite{karamanis2002} or the curvature of the crystal edges (which is neglected in the simulations)~\cite{cabal}. 
A comparison of data and simulation results for radial scans at $d\sim$10~cm (E$_\gamma$=122.1,~1115.5~keV) is shown in Figure~\ref{rad10}. Though the overall fit is good ($\sigma_R$=2.8(3)$\%$), the simulated spectra seems to slightly overestimate the data at higher energies (see Figure~\ref{rad10}). For both the close geometry and distance scans, an excellent agreement is observed between simulations and data. It should be mentioned that the cylindrical symmetry of the crystal was verified with E$_{\gamma}$=59.5~keV by placing the source in all four perpendicular directions close to the detector face.  

\begin{figure}[H]
\centering
\includegraphics[scale=0.6]{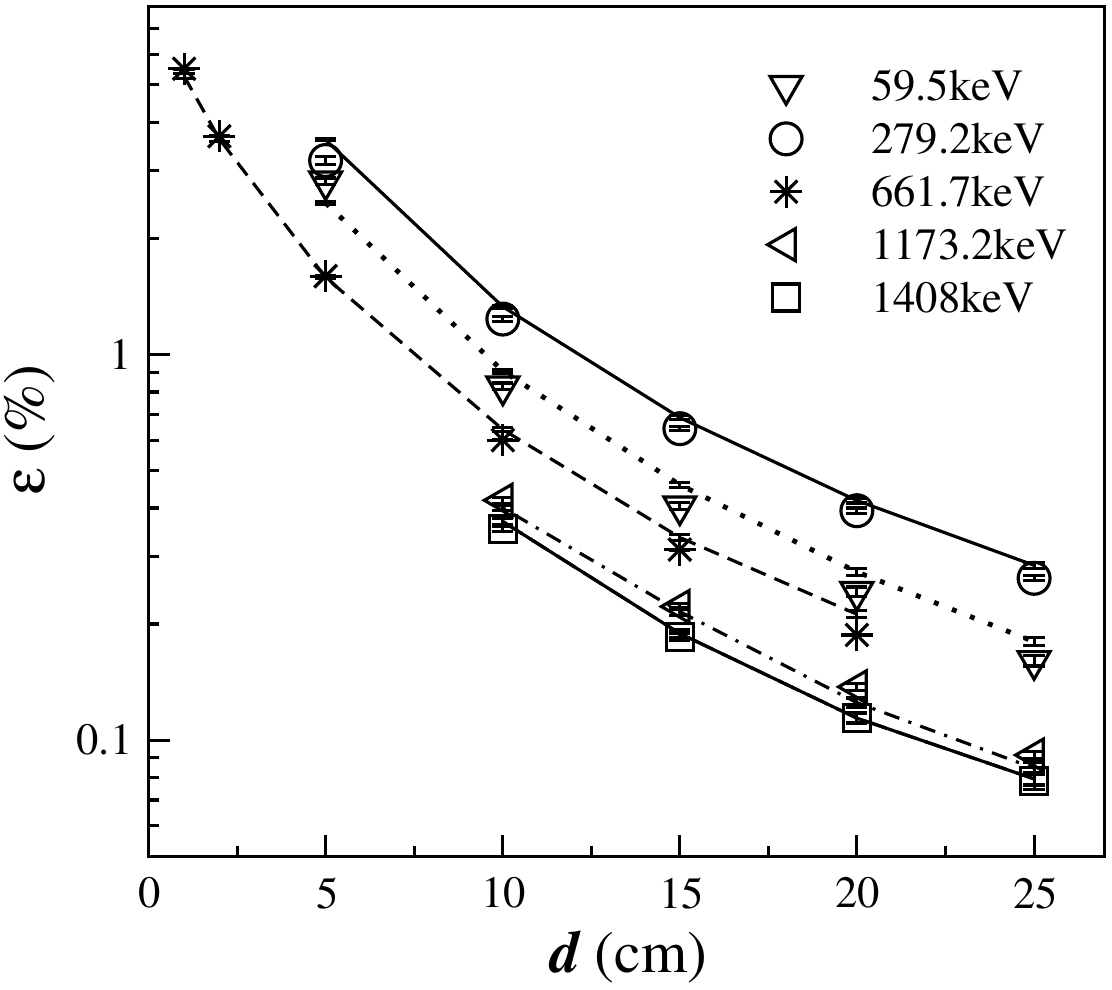} 
\caption{\label{dist10} The $\epsilon^{exp}$ as a function of $d$, distance from the top face of the detector, for different gamma ray energies. Symbols represent the measured data and corresponding $\epsilon^{MC}$ with optimized parameters is shown by lines. }
\end{figure}

\begin{figure}[H]
\centering
\includegraphics[scale=0.6]{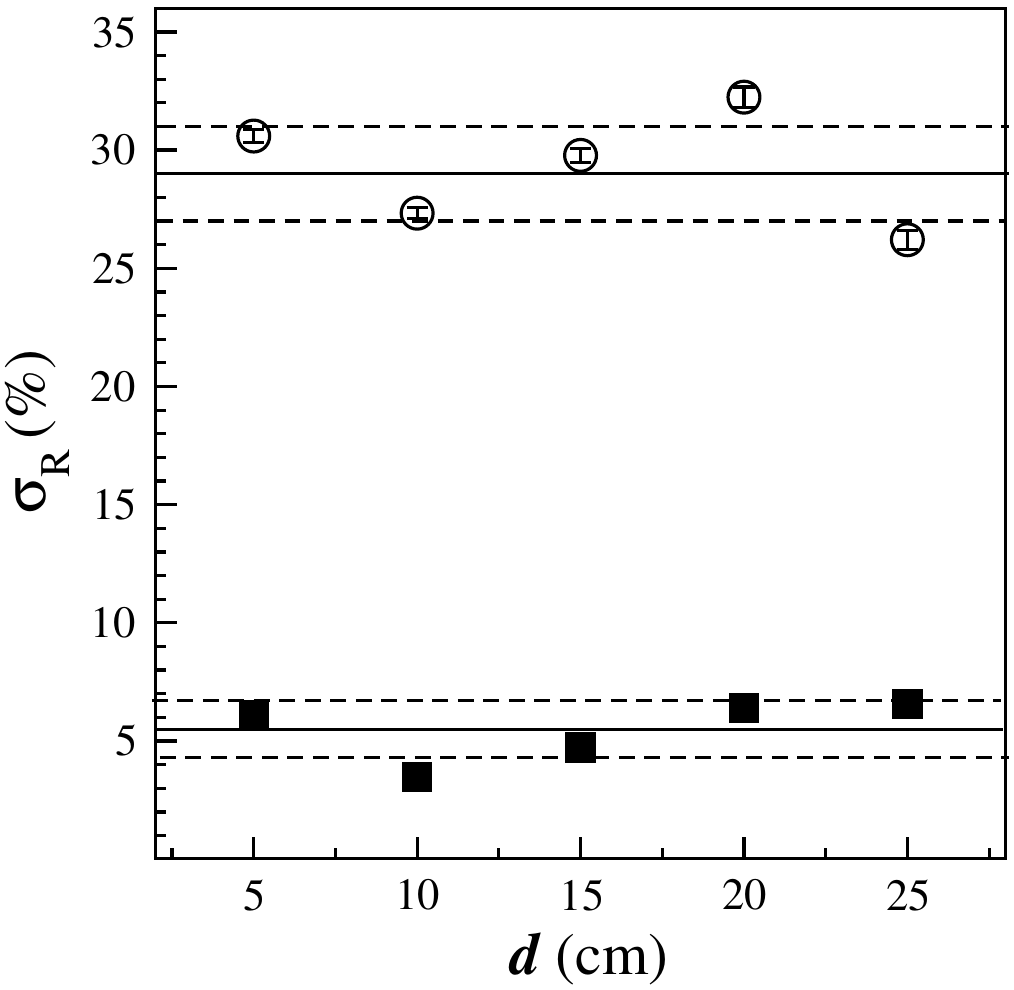} 
\caption{\label{devn} The relative deviation $\rm\sigma_R$ for E$_{\gamma}$=122.1, 279.2, 834.8, 1115.5~keV for $d$=5--25~cm obtained with optimized detector parameters (filled symbols) and with nominal parameters (open symbols). The bold line is the average and the RMS deviation is indicated by dashed lines. Errors are within the point size.}
\end{figure}

The detector model is further tested with distance scan measurements with many sources, E$_\gamma$=59.5, 279.2, 1173.2 and 1408~keV, and results are shown in Figure~\ref{dist10}. 
It is evident from both these figures that the simulations are well able to reproduce the experimental data.  
The effective detector model was used to simulate the volume source geometry (E$_\gamma$=661.7~keV) and results are also plotted in Figure~\ref{dist10}. 
The excellent agreement between measured and simulated values indicate that the optimized model works very well for different source geometries. 
Figure~\ref{devn} displays the relative deviation $\rm\sigma_R$ for $E_{\gamma}$=122.1, 279.2, 834.8, 1115.5~keV as a function of $d$ =5--25~cm. 
It can be seen that the optimized model yields $\rm\sigma_R= 5.46(3)\%$ as opposed to 29.2(3)$\%$ obtained with nominal parameters.
With inclusion of low energy data of E$_\gamma$=59.5~keV, the $\rm\sigma_R$ worsens to $\sim$8.37(4)$\%$.

\begin{figure}[H]
        \begin{minipage}[b]{0.45\linewidth}
          \centering
          \includegraphics[scale=0.36]{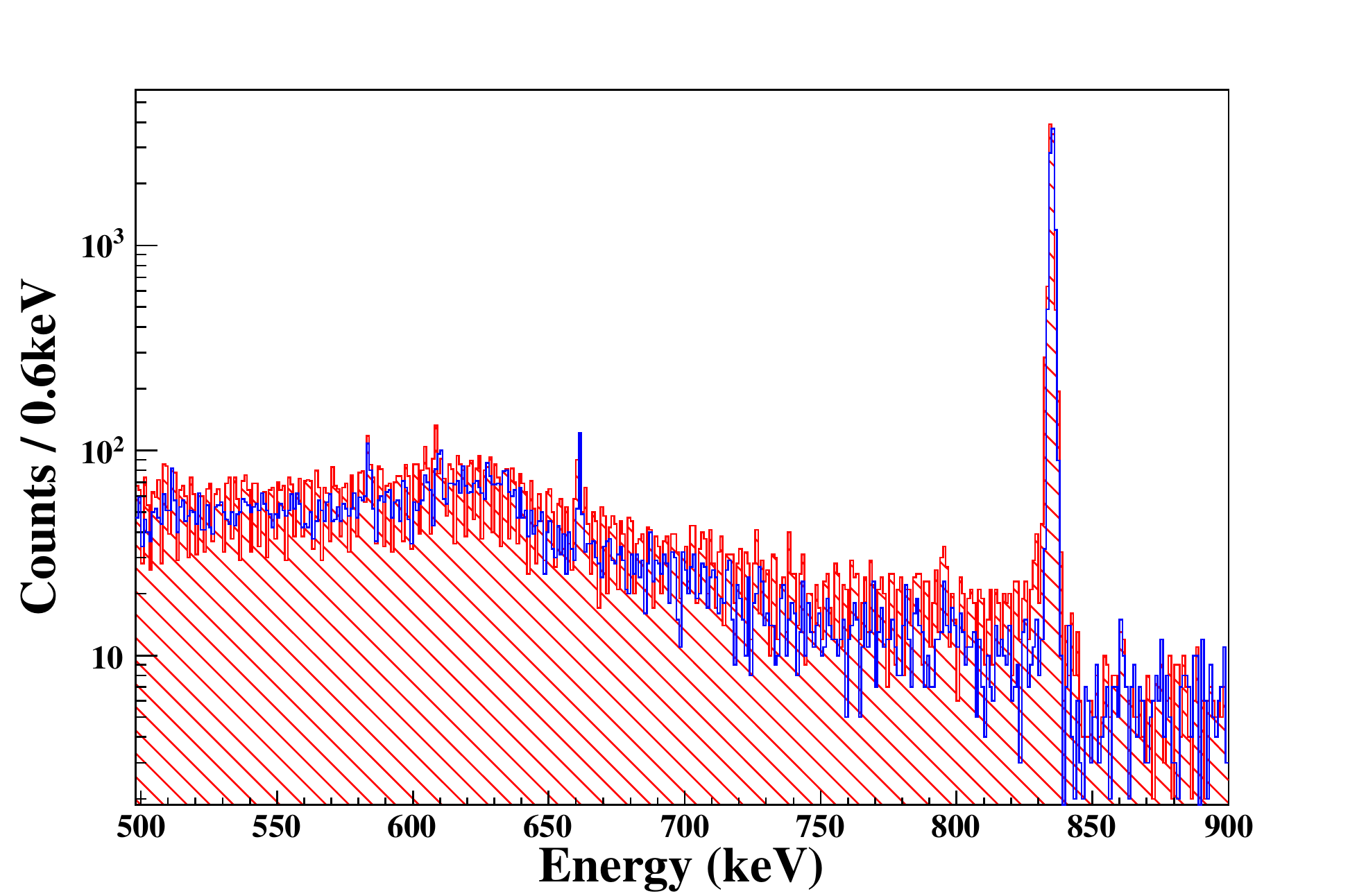}
          \end{minipage}
        \hspace{0.1cm}
        \begin{minipage}[b]{0.45\linewidth}
          \centering
          \includegraphics[scale=0.36]{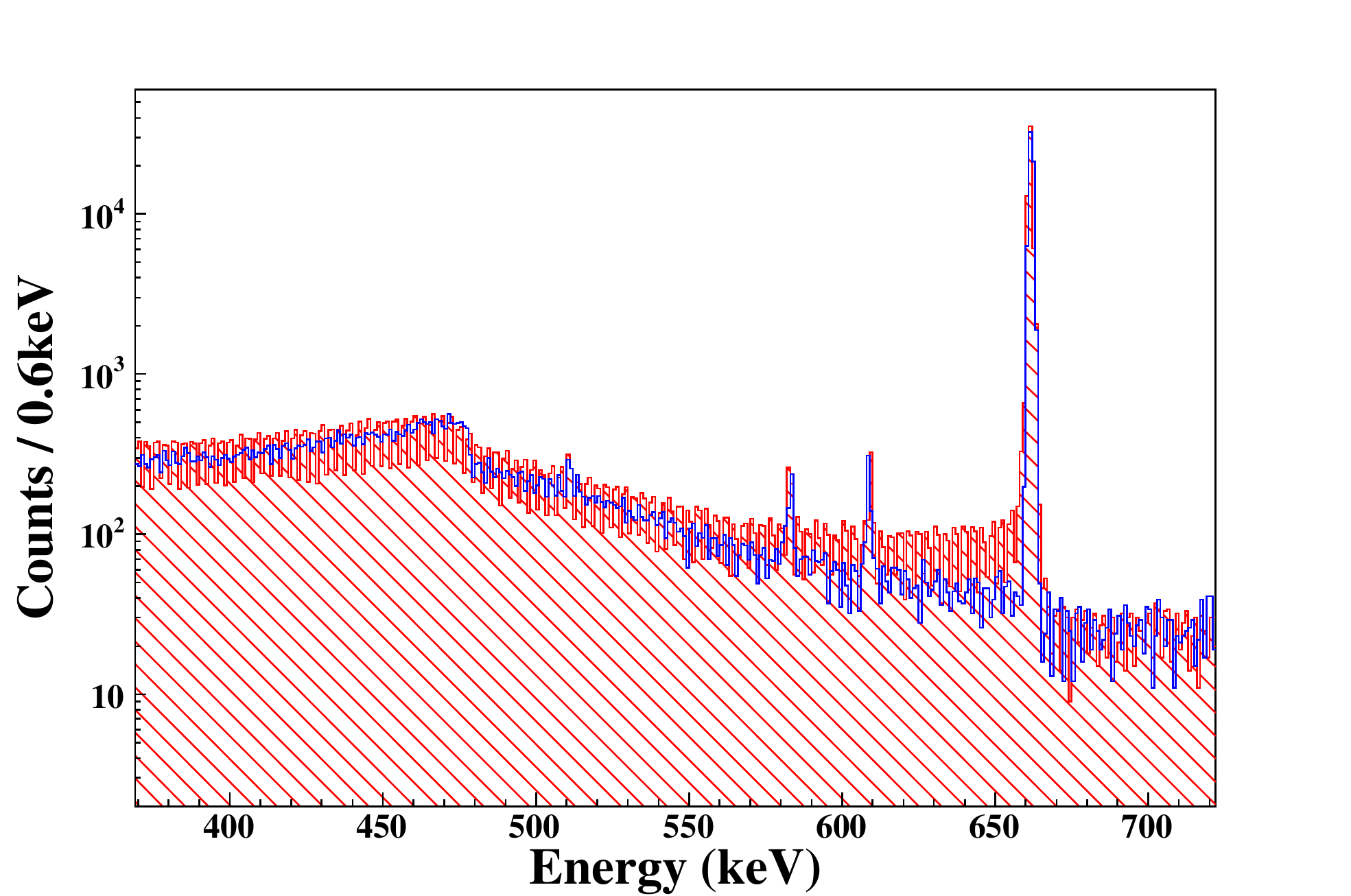}
               \end{minipage}
        \caption{\label{spectra}(Color online) The measured energy spectra (filled red) for $^{54}$Mn extended source (E$_\gamma$=834.8~keV) at $d_s$=25~cm (left panel) and $^{137}$Cs volume source (E$_\gamma$=661.7~keV) at $d$=15~cm (right panel) together with the simulated spectra (blue) after folding in energy resolution of the detector. The room background has been added to the simulated spectrum for comparison. }
      \end{figure}

The measured energy spectra for $^{54}$Mn source (E$_\gamma$=834.8~keV) at $d_s$=25 cm and $^{137}$Cs source (E$_\gamma$=661.7~keV) at $d$=15~cm is shown in Figure~\ref{spectra} together with the simulated spectrum after folding in energy resolution of the detector. 
The room background has been added to the simulated spectrum for comparison with experimental spectrum. 
Even though the detector model was optimized with photopeak efficiency, overall spectral shape including the Compton edge, is very well reproduced. However, a slight low energy tail in the experimental spectrum ($\sim$1.5$\%$) as compared to MC simulations is visible (see Figure~\ref{spectra}).
It should be mentioned that the detector has undergone two thermal cycles and an evacuation during three years of operation without any change in the performance (efficiency and resolution).

\subsection{Low background measurements}
As mentioned in the beginning, the low background counting setup is designed for screening materials for cryogenic bolometer. 
These measurements are usually of long duration (several days) and stability is very important. 
The gain stability of the system is monitored and drifts are found to be negligibly small ($\sim$sub-keV) over a period of several days. 
With a 10~cm thick low activity Pb shield on all sides of the HPGe detector, the background gamma-rays such as E$_{\gamma}$=1460.8~keV ($\rm ^{40}K$) and 
2614.5~keV ($\rm ^{208}Tl$) have been reduced by a factor of $\sim$800(60) and $\sim$200(19), respectively. 
The measured background level of $\rm ^{40}K$ is 51(7) and 166(17) counts per day with and without copper, respectively. Similarly for $\rm^{208}Tl$, 14(2) and 109(14) counts per day are measured in this setup with and without copper, respectively. The background levels can be further improved by addition of cosmic veto and nitrogen flushing.

\begin{figure}[H]
\centering
\includegraphics[scale=0.5]{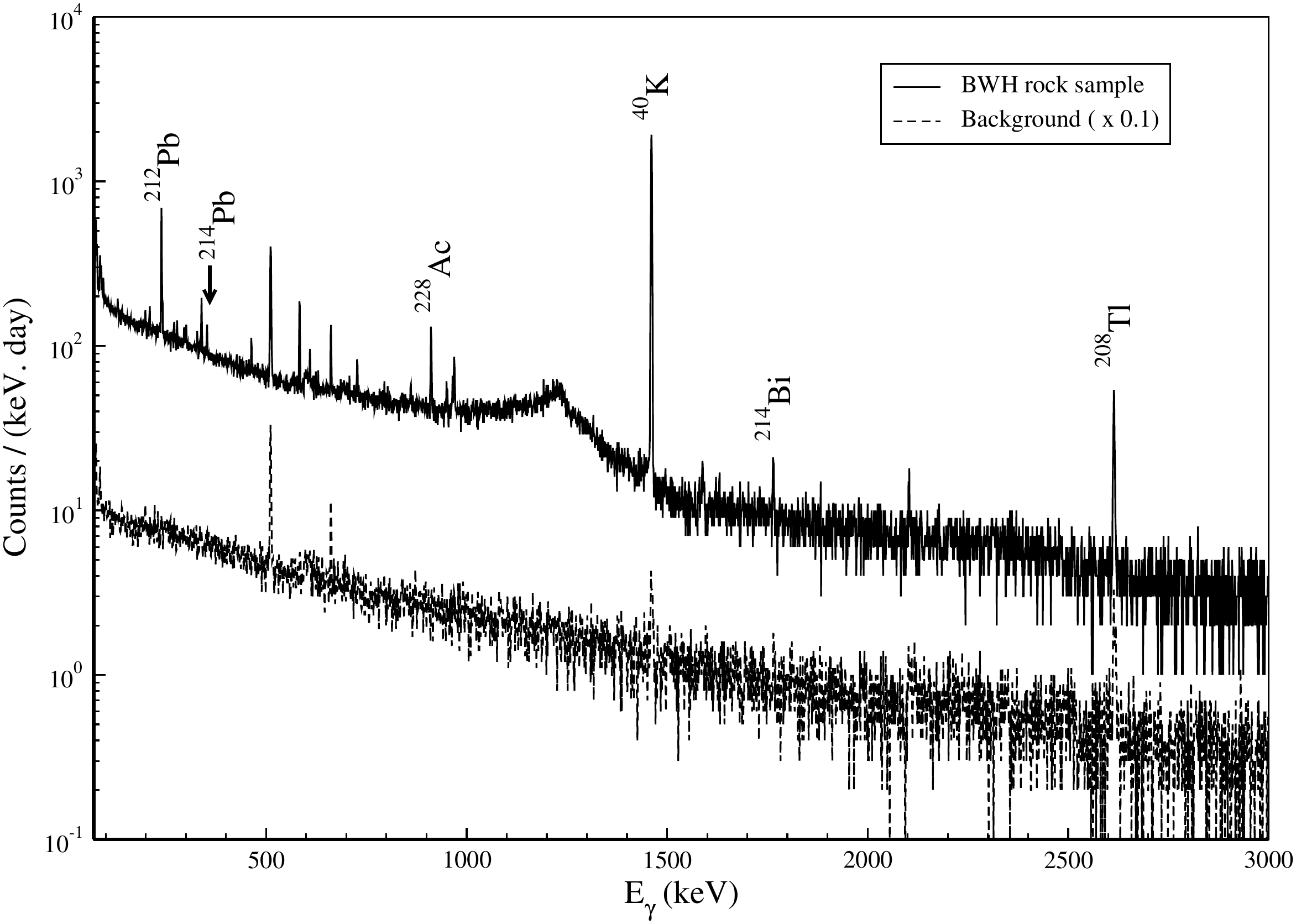} 
\caption{\label{rock}A gamma--ray energy spectrum (bold line) of the rock sample from the INO site (from Bodi West Hills) in the low background setup (only with Pb shield) in a close geometry. The scaled background (dashed line) without the sample is also shown for comparison.}
\end{figure}

The setup has been extensively used to test radio-impurities in various samples like the ETP copper from the Tin bolometer cryostat, $\rm^{nat}$Sn, $^{124}$Sn and sensors etc. The maximum sample size that can be mounted at d$\sim$1~cm is 9~cm $\times$ 9~cm $\times$ 5~cm. The sensitivity of the setup estimated from a sample of copper used in the bolometer setup is about $\sim$1mBq/g for $\rm^{232}Th$ and $\sim$2mBq/g for $\rm^{40}K$. Using this setup, the trace impurity of $\rm^{59}Co$ was estimated to be 1.3(2)~ppb in neutron activated Ge sample~\cite{ntdge}.
In addition, the rock samples from the INO site (Bodi West Hills (BWH))~\cite{rock}, the glass for RPC in ICAL detector~\cite{rpc} have also been studied. 
Figure~\ref{rock} shows a spectrum of the rock sample in a close geometry together with the background spectrum, clearly indicating higher $\rm^{40}K$ content in the sample. 
Table~\ref{bwh} shows estimated concentration of impurities for this sample ($\sim$23~g). 

\begin{table}[H]%
\centering
\caption{Estimated radio-impurity concentrations (N$\rm_x$) in the BWH rock sample from the INO site using low background spectroscopy. }
\label{bwh}
\begin{tabular}{  cc }
\hline
Element& N$\rm_x$ \\ 
       & (mBq/g) \\ \hline
 $^{212}$Pb &11.1(4)  \\ 
 $^{214}$Pb &  1.7(4) \\ 
 $^{228}$Ac&     10.3(7) \\ 
 $^{40}$K   &     1050(16) \\ 
$^{208}$Tl   &     1.8(8) \\ 
$^{214}$Bi   &     7(1) \\\hline
\end{tabular}
\end{table}
It is proposed to study rare events like double beta decay to excited states using this setup, where the efficiency for required source geometry and 
energy range can be obtained using MC simulation technique with the effective detector model.

\section{Conclusions}
A low background counting setup has been made at TIFR consisting of a special HPGe detector surrounded by a low activity copper (5~cm)+lead (10~cm) shield. 
Detailed measurements are performed with point and extended geometry sources to generate an effective model of the detector with GEANT4 based Monte Carlo simulations. 
The active volume obtained is about 20$\%$ smaller than the nominal value supplied by the manufacturer. 
The effective detector model agrees  within 5.46(3)$\%$ with experimental data over a wide energy range of 100--1500~keV.
Using the simulated efficiencies, impurities at ppb level in various samples have been measured. 
This low background counting setup will be used for qualification and selection of radio-pure materials to be used in the prototype bolometer R$\&$D and for rare event studies.

\section{Acknowledgements}
The authors would like to thank Mr.~M.S.~Pose and Mr.~K.V.~Divekar for help during the setup.


\begin{thebibliography}{00}

\bibliographystyle{elsarticle-num}

\bibitem{leonard} D.S.~Leonard {\it et. al.}, {Nucl. Instr. and Meth. A} {\bf591} (2008) 490.

\bibitem{arnold} R.~Arnold {\it et. al.}, {Nucl. Instr. and Meth. A} {\bf354} (1995) 338.

\bibitem{agostiniprl}M.~Agostini {\it et. al.}, {Phys. Rev. Lett.} {\bf 111} (2013) 122503.

\bibitem{auger}M.~Auger {\it et. al.}, {Phys. Rev. Lett.} {\bf 109} (2012) 032505.

\bibitem{agostini}M.~Agostini {\it et. al.}, {arXiv:1306.5084}.

\bibitem{bellini}F.~Bellini, C.~Bucci, S.~Capelli, O.~Cremonesi, L.~Gironi, M.~Martinez, M.~Pavan, C.~Tomei , M.~Vignati, { Astroparticle Physics} {\bf 33} (2010) 169.

\bibitem{andreotti}E.~Andreotti {\it et al.}, Astroparticle Physics {\bf34} (2010) 18.

\bibitem{argyriades} J.~Argyriades {\it et. al.}, {Nucl. Instr. and Meth. A} {\bf606} (2009) 449.

\bibitem{budjas}D.~Budjas, M.~Heisel, W.~Maneschg, H.~Simgen, Appl. Radiat. and Isot. {\bf67} (2009) 706.

\bibitem{cabal}Fatima Padilla Cabal, Neivy Lopez-Pino, Jose Luis Bernal-Castillo, Yisel Martinez-Palenzuela, Jimmy Aguilar-Mena, Katia D'Alessandro, Yuniesky Arbelo, Yasser Corrales, Oscar Diaz, Appl. Radiat. and Isot. {\bf68} (2010) 2403.

\bibitem{diaz}N.~Cornejo~Diaz, M.~Jurado~Vargas, {Nucl. Instr. and Meth. A} {\bf586} (2008) 204.

\bibitem{hardy}J.C.~Hardy, V.E.~Iacob, M.~Sanchez-Vega, R.T.~Effinger, P.~Lipnik, V.E.~Mayes, D.K.~Willis, R.G.~Helmer, Appl. Radiat. and Isot. {\bf 56} (2002) 65.

\bibitem{helmer}R.G.~Helmer, J.C.~Hardy, V.E.~Iacob, M.~Sanchez-Vega, R.G.~Neilson, J.~Nelson, {Nucl. Instr. and Meth. A} {\bf  511} (2003) 360.

\bibitem{hernandez}F.~Hernandez, F.~El-Daoushy, {Nucl. Instr. and Meth. A} {\bf  498} (2003) 340.

\bibitem{hurtado}S.~Hurtado, M.~Garcia-Leon, R.~Garcia-Tenorio, {Nucl. Instr. and Meth. A} {\bf518} (2004) 764.

\bibitem{karamanis2003}D.~Karamanis, {Nucl. Instr. and Meth. A} {\bf  505} (2003) 282.

\bibitem{Raina}P.K.~Raina {\it et. al.}, Ed. V.K.B.~Kota and U.~Sarkar, Narosa Publishers (2007).

\bibitem{VSingh}V.~Nanal, International Nuclear Physics Conference: 2013, EPJ Web of Conferences ({\it in press}).

\bibitem{pramana}V.~Singh, S.~Mathimalar, N.~Dokania, V.~Nanal, R.G.~Pillay, S.~Ramakrishnan, Pramana {\bf 81} (2013) 719.

\bibitem{ino}N.K.~Mondal, Pramana {\bf 79} (2012) 1003.

\bibitem{qvalue}D.A.~Nesterenko {\it et. al.}, Phys. Rev. C {\bf86} (2012) 044313.

\bibitem{barabash2} A.S.~Barabash, Ph.~Hubert, A.~Nachab, S.I.~Konovalov, I.A.~Vanyushin, V.~Umatov, Nucl. Phys. A {\bf807} (2008) 269.

\bibitem{belli} P.~Belli {\it et. al.}, Phys. Rev. C {\bf83} (2011) 034603.

\bibitem{belli1} P.~Belli {\it et. al.}, Nuclear Physics A {\bf846} (2010) 143.

\bibitem{belli2} P.~Belli {\it et. al.}, Nuclear Physics A {\bf859} (2011) 126.

\bibitem{barabash3} A.~Barabash {\it et. al.}, AIP Conf. Proc. {\bf1417} (2011) 28.

\bibitem{belli3} P.~Belli {\it et. al.}, Phys. Rev. C {\bf87} (2013) 034607.

\bibitem{boson}Jonas~Boson, Goran~Agren, Lennart~Johansson, {Nucl. Instr. and Meth. A} {\bf587} (2008) 304. 

\bibitem{lamps}http://www.tifr.res.in/$\sim$pell/lamps.html

\bibitem{geant}S.~Agostinelli {\it {et. al.}}, { Nucl. Instr. and Meth. A} {\bf506} (2003) 250.

\bibitem{root}Rene~Brun, Fons~Rademakers, Nucl. Instr. and Meth. A {\bf 389} (1997) 81.

\bibitem{knoll}G.F.~Knoll, Radiation Detection and Measurement, third ed., Wiley, New York, (2000).

\bibitem{rodenas}J.~Rodenas, A.~Pascual, I.~Zarza, V.~Serradell, J.~Ortiz, L.~Ballesteros, { Nucl. Instr. and Meth. A} {\bf496} (2003) 390.

\bibitem{huy}N.Q.~Huy, D.Q. Binh, V.X. An, Nucl. Instr. and Meth. A 573 (2007) 384.

\bibitem{karamanis2002}D.~Karamanis, V.~Lacoste, S.~Andriamonje, G.~Barreau, M.~Petit, {Nucl. Instr. and Meth. A} {\bf487} (2002) 477.

\bibitem{ntdge} N.~Dokania {\it {et. al.}}, DAE Symp. on Nucl. Phys. {\bf 56} (2011) 1136.

\bibitem{rock} N.~Dokania {\it {et. al.}}, DAE Symp. on Nucl. Phys. {\bf 56} (2011) 1138.

\bibitem{rpc} V.M.~Datar, Satyajit~Jena, S.D.~Kalmani, N.K.~Mondal, P.~Nagaraj, L.V.~Reddy, M.~Saraf, B.~Satyanarayana, R.R.~Shinde, P.~Verma, Nucl. Instr. and Meth. A {\bf602} (2009) 744.

\end{thebibliography}
\end{document}